\documentclass[aps,prl,twocolumn,groupedaddress,epsfig,amsmath,floatfix,showpacs]{revtex4}
\usepackage{epsfig,amsmath}
\usepackage{graphicx}
\usepackage{dcolumn}
\usepackage{bm}

\begin{document}

\title{In-plane Theory of Non-Sequential Triple Ionization}

\author{Phay J. Ho and J. H. Eberly}
\affiliation{Department of Physics and Astronomy, University of 
Rochester, Rochester NY 14627, USA}
\email{phayho@pas.rochester.edu}

\begin{abstract}%
We describe first-principles in-plane calculations of non-sequential triple ionization (NSTI) of atoms in a linearly polarized intense laser pulse. In a fully classically correlated description, all three electrons respond dynamically to the nuclear attraction, the pairwise e-e repulsions and the laser force throughout the duration of a 780nm laser pulse.  Nonsequential ejection is shown to occur in a multi-electron, possibly multi-cycle and multi-dimensional, rescattering sequence that is coordinated by a number of sharp transverse recollimation impacts.
\end{abstract}


\maketitle

Modern high-power Ti:sapphire laser systems can now easily generate high repetition-rate femtosecond laser pulses with intensities greater than $10^{13}$ W/cm$^2$ or as high as $10^{20}$ W/cm$^2$.  In this range of intensities, atomic and molecular binding potentials are so strongly distorted that many electrons can be removed. Recently, using such laser pulses, the production rate of almost all charged states of Xe, up to Xe$^{+20}$, have been examined \cite{Yamakawa-etal04}.  This initiative and several other recent experiments \cite{N-electronexpts} are part of a continuing effort to understand how laser energy is coupled to atomic and molecular targets.

The emergence of these new experimental efforts has made the theoretical study of multi-electron ejection from atoms and molecules necessary.  Experience from studies on double-electron ejection \cite{NSDI-atoms, NSDI-molecules,Corkum, smatrix,coulomb-focusing, softcollision, Texas, weber_He, moshammer_Ne, Panfili-etal02, Haan-etal02, Ho-etal05} supports the generally accepted recollision picture \cite{Corkum}, in which a leading electron is liberated from the target by the laser field and is driven back by the field to deposit the energy that it gains from the field to the second electron \cite{NSDIrecollision}.

The inclusion of fully time-dependent pairwise e-e interactions poses a daunting challenge for any quantum-theoretical investigation going beyond the single-pair treatment by the Taylor group \cite{Taylor} for helium. There is probably no realistic prospect for extending the quantum mechanical treatment to the multi-pair interactions participating in triple ionization under the relevant conditions, i.e., for laser wavelengths and under highly nonperturbative fields with femtosecond and shorter time resolution.  Thus, a different theoretical strategy is called for.

It has been shown for double ionization that very large one-dimensional \cite{Panfili-etal01} and three-dimensional \cite{Haan-etal04} classical ensemble calculations  have been in conformity with  observed double-ionization phenomena (also see \cite{Ho-etal05} and references therein). In this note we show that this classical approach can be substantially extended, and we report the first results of a fully dynamical treatment of ionization of three electrons in a two-dimensional reaction plane. Our extension allows (a) tracking of three distinct electron pairs, (b) clear interpretation of transverse motion for a recolliding electron, and (c) inclusion of structural features keyed to specific atomic targets. The first extension is  necessary for any first-principles treatment of NSTI, and the second exposes the role of weak but sharp transverse impacts that guide the recolliding electron toward its NSTI end game. The third extension opens a domain for study that is too large for the available space here and its results will be reported elsewhere \cite{Ho-Eberly06}. Altogether, we believe that our results present the first easily interpreted and detailed view of NSTI.

\begin{figure}[h]
\centerline{\includegraphics[width=2in]{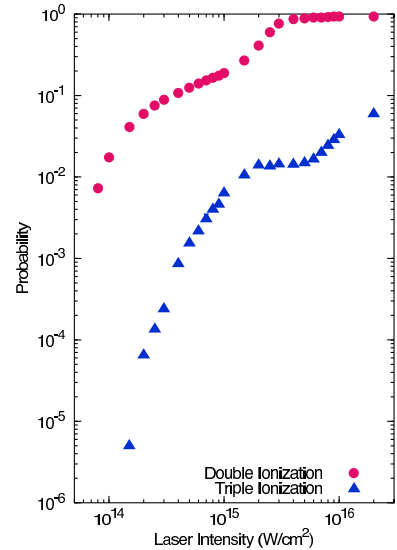} }
\caption{(Color Online)The so-called knee signature of non-sequential double ionization is shown for triple ionization along with the corresponding knee obtained earlier for the 2+ ion count. Both ion-count curves were obtained from the fully classical model of strong-field ionization familiar from the literature (see \cite{Javanainen-etal87}).}
\label{fig.knee}
\end{figure}

\begin{figure*}
\centerline{\includegraphics[width=7in]{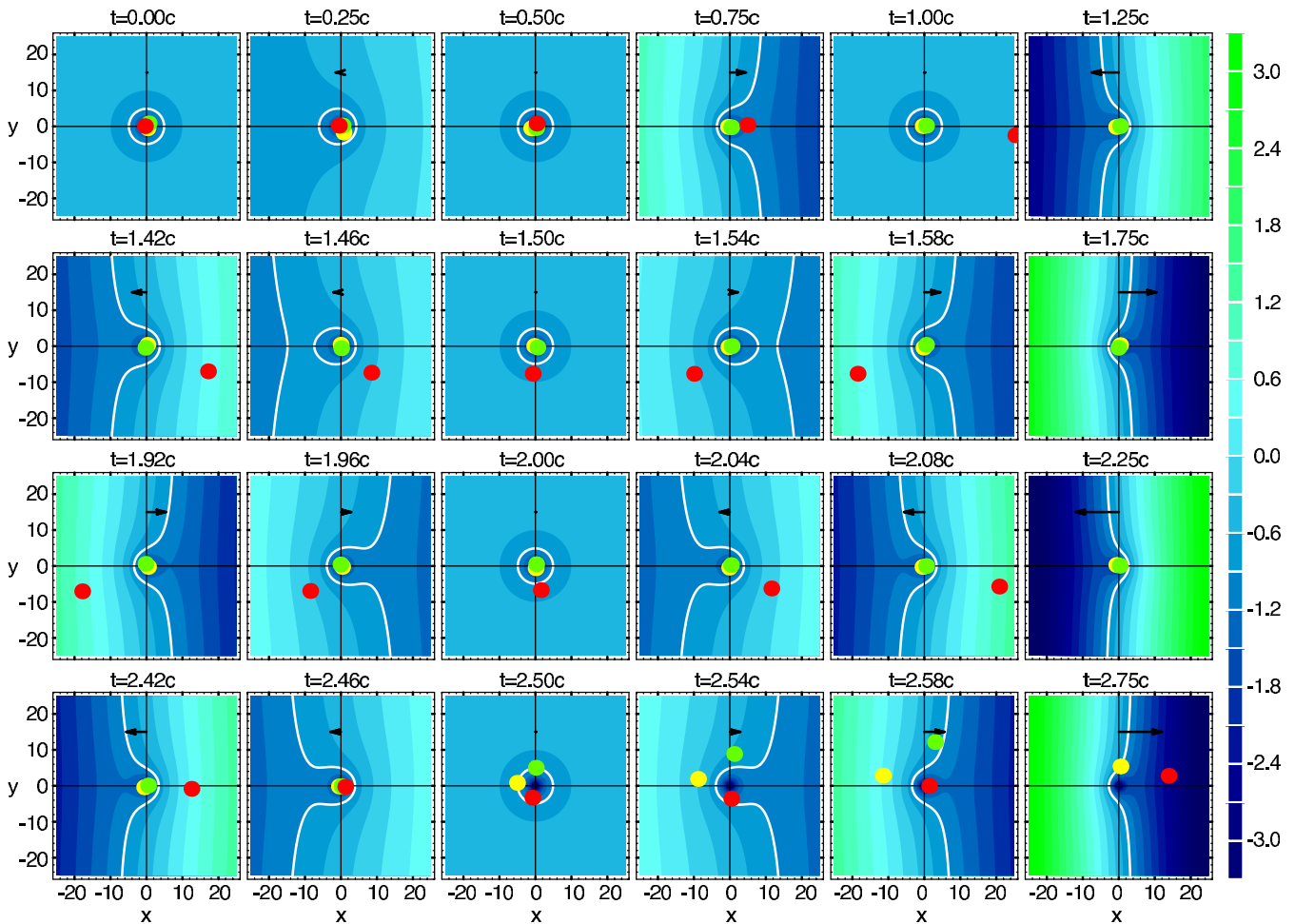}
              }
\caption{(Color) Chronological sequence of a planar triple ionization event. The three colored balls represent electrons in a plane that indicates their potential energies (due to laser and nuclear forces) by tipping back and forth every half laser cycle. Blue shading indicates the lowered side.  The axes show the longitudinal (x, along laser polarization) and transverse (y) locations of the electrons in atomic units.  The energy contours have a difference of 0.3 a.u., as indicated by a vertical legend on the right.  For reference, the contour line with value of -0.6 a.u. is highlighted with a white line.  The horizontal arrow in each plot shows the direction and strength of the laser force in that snapshot.  The first row shows an electron escaping to the continuum, and the next three rows show three suscessive returns of that electron. Note that groups of time intervals are selected to capture key interaction events and are not equally spaced.}
\label{fig.ti}
\end{figure*}

First we show in Fig.~\ref{fig.knee} that NSTI exists in the conventional sense, i.e., that a ``knee" is predicted for the triple-ion count. Second, Fig. \ref{fig.ti} shows a sequence of snapshots running from t=0 throughout an NSTI episode, demonstrating the nature of NSTI ejection. Third, we will show that a time-sequenced analysis of the transverse component of motion gives new insight into the synergy between longitudinal and transverse forces. It reveals unexpectedly sharp transverse impacts that are needed to create near-axial alignment of the recolliding electron, which allows it to strike the core and free the other two electrons.

Fig.~\ref{fig.ti} shows the motion of all three electrons in the reaction  plane during evolution in a laser pulse with peak intensity of $I=8.0 \times 10^{14}$W/cm$^2$.  In this sequence, one of the bound electrons (red) first leaves the nucleus and returns with a transverse displacement taking it outside the -0.6 a.u. marker.  This electron then returns to the core three times and after twice missing the bound electrons it knocks both out in the third pass.  This pathway typifies the rescattering scenario found among the NSTI trajectories obtained in such a laser pulse.

Contours and shading in Fig. \ref{fig.ti} show the combined potential energies from laser and nuclear forces on the individual electrons. The tipping back and forth of the plane shows how the binding potential centered at $x=y=0$ is deformed by the laser field. Barrier depression by the laser allows one of the initially bound electrons to leave the nucleus without tunneling.  The departure time corresponds to the phase of maximum field strength in the second half of the laser cycle because then the maximum degree of barrier depression is sufficient. The process of barrier depression at $I=8.0 \times 10^{14}$W/cm$^2$ has little or no effect on the remaining two electrons.

This is the first view of different NSTI stages from a first-principles fully dynamical and fully correlated (albeit classical) calculation. Theoretical exploration of intense-field triple ionization under experimental conditions so far is very limited.  Sacha and Eckhardt \cite{Sacha-Eckhardt-triple} have calculated the momentum distribution of a triply charged ion starting from a highly excited compound state, where the three electrons need not be bound initially.  Liu, et al.,  \cite{LiuX-etal06} have presented the results of an energy-sharing thermalization Ansatz following tunneling initiation. In contrast to these two approaches, our calculation starts with all 3 electrons bound as the calculation begins.

Details of our method have been explained before \cite{Javanainen-etal87, MethodDetails}, but the elements are straightforward. Under the influence of a laser pulse, the responses of each 3-e trajectory in both the longitudinal ($x$) and tranverse ($y$) directions are given by solutions of the nonlinear Newtonian ode's:
\begin{eqnarray}
\frac{d^2 x_i}{dt^2}&=&-E(t)-\frac{3x_i}{(x_i^2+y_i^2+a^2)^{3/2}} \nonumber \\ 
&&+\sum_{j\ne i}\frac{(x_i-x_j)}{((x_i-x_j)^2+(y_i-y_j)^2+b^2)^{3/2}},\label{eqn.ode1}
\end{eqnarray}
\begin{eqnarray}
\frac{d^2 y_i}{dt^2}&=&-\frac{3y_i}{(x_i^2+y_i^2+a^2)^{3/2}} \nonumber \\ 
&& +\sum_{j\ne i}\frac{(y_i-y_j)}{((x_i-x_j)^2+(y_i-y_j)^2+b^2)^{3/2}},\label{eqn.ode2} 
\end{eqnarray}
where the subscript $i$ is the electron label that runs from 1 to 3, and $E(t)$ is our 20fs 780nm laser pulse, which has a trapezoidal pulse shape with a 2-cycle turn-on, a 4-cycle plateau and a 2-cycle turn-off.    About 4\% of the trajectories are found to be triply ionized at the end of the pulse.

\begin{figure}[t]
\centerline{\includegraphics[height=1.8in]{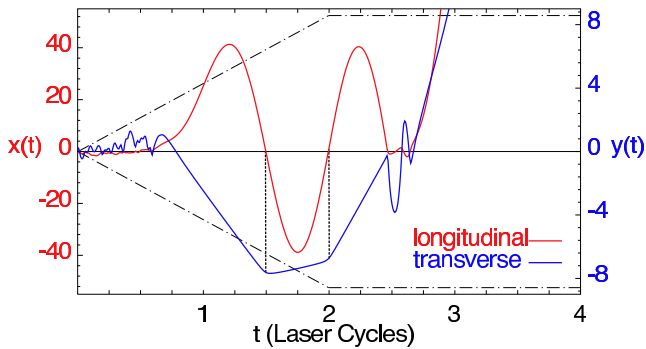}
              }
\caption{(Color Online) Longitudinal and transverse displacement of 
the rescattering electron.  The light (red) and dark (blue) lines 
plot the longitudinal (left scale) and transverse (right scale) 
displacement.  The vertical dotted lines indicate first two return 
times of the rescattering electron, i.e. when the red line crosses 
x=0, and divide the growth of the transverse spread into three 
regimes.  The dashed-dotted line is a marker for the laser field 
envelope.  The recollision stage takes place from rougly t=0.750c to 
about t=2.500c.}
\label{fig.trvslgx}
\end{figure}

In addition to being driven by the laser, the recolliding electron also undergoes transverse displacement as seen in the first two returns in Fig. ~\ref{fig.ti}.  The nuclear attraction in Eqs. (\ref{eqn.ode1}) and (\ref{eqn.ode2}), of course, provides the transverse force, and Coulomb focusing takes place, but analysis of the recollision scenario shows that this focusing has dominant features not previously demonstrated but easily explained. The laser force carries the recolliding electron so far from the nucleus that whatever transverse motion it has is simply continued at a constant velocity during its longitudinal excursions. That is, the electron drifts freely transversely except when it is very close to $x=0$, at which time the nucleus can be effective, but only very briefly. Thus in Fig. ~\ref{fig.trvslgx} we see three long sequences of linear transverse drift, first away from the nucleus and then toward  it, rather than a steady focusing. One clearly sees that the changes in transverse drift motion occur abruptly just at the times the recolliding electron crosses the nucleus ($x=0$).

\begin{figure}[b]
\centerline{\includegraphics[width=3.2in]{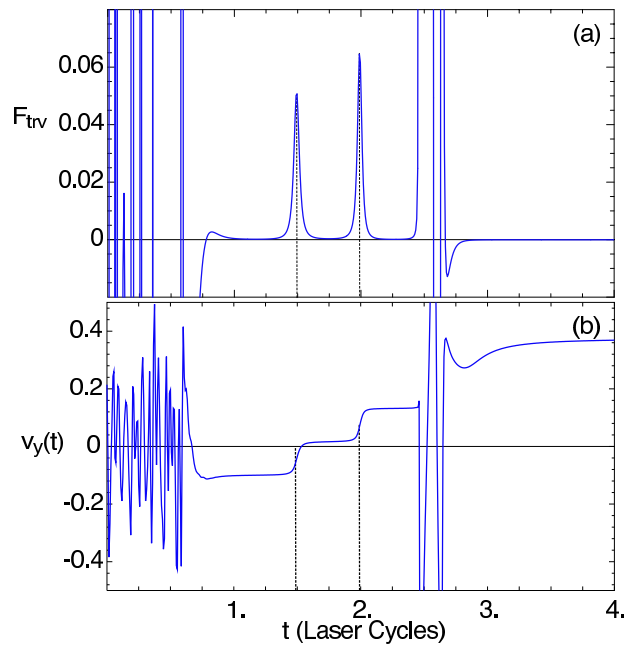}
              }
\caption{(Color Online) Transverse nuclear force (top) and transverse 
momentum (bottom) of the rescattering electron.  The full range of 
the these quantities is beyond what are plotted.  The two sharp 
impulses and two step-like jumps in the top and bottom panels 
respectively coincide with the first two return times, which are 
denoted by two vertical dotted lines.}
\label{fig.trvslgv}
\end{figure}

The imprint of the longitudinal laser-driven motion can be identified clearly in the transverse motion.  Rather than a continuous pull, the nuclear force affects the electron as a series of impulses as shown in Fig. \ref{fig.trvslgv}a.  These impulses produce step-like jumps in the transverse momentum (see Fig. \ref{fig.trvslgv}b).  In our trajectory, the first impulse has reversed the direction of transverse momentum so that the electron runs towards the nucleus at a constant transverse speed instead of moving away.  The second impulse merely raises this speed of reaching the nucleus.

In summary, we have used a very large ensemble of classical trajectories for three electrons simultaneously to examine high-intensity triple ionization, within a planar reaction scenario. The benefit of the classical approach is that it permits trajectories to be followed deterministically and displayed graphically for interpretation. Our calculation shows that a laser driven multi-electron correlation drives the triple ionization. This generates a triply charged ion count vs. intensity curve, which displays a three-electron non-sequential ``knee" signature as seen in Fig. \ref{fig.knee}.

The recolliding electron scenario of Fig. \ref{fig.ti} is shown to lead to a 3-electron complex (last row of the figure) just prior to the coordinated NSTI event. We suggest that this might be interpreted as showing an attosecond-scale ``thermalization", consistent with the very recent prediction of Liu, et al. \cite{LiuX-etal06}.  A key feature, not previously available, is the detailed effect of the nuclear force as modulated by the laser, producing a series of weak sharp impulses on the recolliding electron.  The formation of these impulses is the basis for the transverse re-collimation and for setting up the final ionization reaction itself in the intense low frequency field, a sequence evident in Fig. \ref{fig.trvslgx}.

This work was supported by DOE Grant DE-FG02-05ER15713.


\begin{thebibliography}{99}
\bibitem{MethodDetails} We use the so called Rochester potential \cite{Javanainen-etal87}, $V(r) = Z_1 Z_2 / \sqrt{r^2+\alpha^2}$, where $Z_1$ and $Z_2$ are the charges of the nucleus and an electron, $r$ is the distance between these two particles and $\alpha$ is the screening length, which need not be the same for the nuclear binding potential and the e-e repulsion potential.  In our calculations, values of the nuclear screening length, $a=1.00$, and the e-e repulsion screening length, $b=0.25$, are used to avoid auto-ionization of one or more electrons in the absence of a laser field \cite{Ho-Eberly06} and to give the additional benefit of a higher yield of triple-ionization events.  Our three-electron system is represented by a micro-canonical ensemble with up to 1 million trajectories, each with a fixed initial energy of -4.63 a.u. and zero total angular momentum.  Each electron however needs not have zero angular momentum.  To populate this ensemble, we use the pilot atom method \cite{Panfili-etal02}, which automatically provides an appropriate statistical weighting of the phase space of this 3-e system that satisfies our energy and angular momentum conditions.
\end{thebibliography}
\end{document}